\documentclass[times,twocolumn,review,preprint]{elsarticle}

\usepackage{jasr}
\usepackage{framed,multirow}

\usepackage{amssymb,amsmath}
\usepackage{latexsym}

\usepackage{colortbl}

\usepackage[switch]{lineno}

\usepackage{url}
\usepackage{xspace,tabularx}
\usepackage[dvipsnames,table]{xcolor}

\usepackage{xcolor}
\usepackage[colorlinks=true]{hyperref}
\AtBeginDocument{\hypersetup{citecolor=black!40!green,linkcolor=black}} %
\usepackage[noabbrev]{cleveref}

\usepackage{glossaries}
\glsdisablehyper
\usepackage[
	per-mode=symbol,
	separate-uncertainty=true,
	list-units = single,%
    range-units = single,%
    multi-part-units = single,%
   ]{siunitx}

\newacronym{GRACE}{GRACE}{Gravity Recovery And Climate Experiment}
\newacronym{GFO}{GRACE-FO}{Gravity Recovery And Climate Experiment -- Follow On}
\newacronym{LRI}{LRI}{Laser Ranging Interferometer}
\newacronym{KBR}{KBR}{K-Band Ranging}
\newacronym{MWI}{MWI}{Microwave Instrument}
\newacronym{LRP}{LRP}{Laser Ranging Processor}
\newacronym{SEU}{SEU}{Single Event Upset}
\newacronym{LEO}{LEO}{Low Earth Orbit}
\newacronym{SAA}{SAA}{South-Atlantic Anomaly}
\newacronym{GPS}{GPS}{Global Positioning System}
\newacronym{LISA}{LISA}{Laser Interferometer Space Antenna}
\newacronym{QPD}{QPD}{Quadrant Photodiode}
\newacronym{DWS}{DWS}{Differential Wavefront Sensing}
\newacronym{FPGA}{FPGA}{Field Programmable Gate Array}
\newacronym{FIR}{FIR}{Finite Impulse Response}
\newacronym{LUT}{LUT}{Look-Up Table}
\newacronym{rms}{rms}{root-mean-square}
\newacronym{ASD}{ASD}{amplitude spectral density}
\newacronym{PSD}{PSD}{power spectral density}
\newacronym{TID}{TID}{Total Ionizing Dose}
\newacronym{SEE}{SEE}{Single Event Effects}
\newacronym{SEL}{SEL}{Single Event Latchup}
\newacronym{JPL}{JPL}{Jet Propulsion Laboratory}

\newcommand{\rtHz}{\sqrt{\mathrm{\hertz}}}

\newcommand{\GFC}{\mbox{GF-1}\xspace}
\newcommand{\GFD}{\mbox{GF-2}\xspace}
\newcommand{\F}[1]{\ensuremath{F_{\!#1}\xspace}}
\newcommand{\D}[1]{\ensuremath{D_{#1}\xspace}}
\newcommand{\LUT}[3]{\ensuremath{\mathrm{LUT}_{\mathrm{#1}}^{#2,#3}\xspace}}
\newcommand{\pdfunc}[3][]{\ensuremath{\mathcal{L}^{#1}\!\left(#2\,|\,#3\right)}}
\newcommand{\loglikeli}[3][]{\ensuremath{\ell^{#1}\!\left(#2\,|\,#3\right)}}

\newcommand{\transpose}{\mathsf{T}}

\definecolor{pale_green}{RGB}{204,221,170}
\definecolor{pale_yellow}{RGB}{238,238,187}
\definecolor{pale_red}{RGB}{255,204,204}
\definecolor{pale_gray}{RGB}{221,221,221}

\journal{Advances in Space Research}
\begin{document}

\verso{Malte Misfeldt \textit{et. al.}}

\begin{frontmatter}

\title{Disturbances from Single Event Upsets in the GRACE Follow-On Laser Ranging Interferometer}%

\author[1,2]{Malte \snm{Misfeldt}\corref{cor1}}
\ead{malte.misfeldt@aei.mpg.de}
\cortext[cor1]{Corresponding author:}
\author[1,2]{Pallavi \snm{Bekal}}
\author[1,2]{Vitali \snm{Müller}}
\author[1,2]{Gerhard \snm{Heinzel}}

\address[1]{Max-Planck Institute for Gravitational Physics, Callinstraße 38, 30167 Hannover, Germany}
\address[2]{Institut für Gravitationsphysik, Leibniz Universität Hannover, Callinstraße 38, 30167 Hannover, Germany}

\received{}
\finalform{}
\accepted{}
\availableonline{}
\communicated{}

\begin{abstract}
The \gls{GFO} satellite mission (2018-now) hosts the novel \gls{LRI}, a technology demonstrator for proving the feasibility of laser interferometry for inter-satellite ranging measurements. The \gls{GFO} mission extends the valuable climate data record of changing mass distribution in the system Earth, which was started by the original \glsunset{GRACE}\gls{GRACE} mission (2002-2017). 
The mass distribution can be deduced from observing changes in the distance of two low-earth orbiters employing interferometry of electromagnetic waves in the K-Band for the conventional \gls{KBR} and in near-infrared for the novel \gls{LRI}. 

This paper identifies possible radiation-induced \gls{SEU} events in the \gls{LRI} phase measurement. We simulate the phase data processing within the \gls{LRP} and use a template-based fitting approach to determine the parameters of the \gls{SEU} and subtract the events from the ranging data. Over four years of \gls{LRI} data, 29 of such events were identified and characterized.
\end{abstract}


\end{frontmatter}

\glsresetall
\section{Introduction}
\label{sec::introduction}
The ongoing \gls{GFO} space mission consists of two nearly identical formation-flying satellites, launched on May 22nd, 2018 \citep{Kornfeld2019}. The objective of \gls{GFO} and its predecessor \glsunset{GRACE}\gls{GRACE} (2002-2017) is to study mass redistribution within the Earth system by observing the differential gravimetric pull on the satellites \citep{Wahr1998}. Both satellites share a similar polar orbit at an altitude of about \SI{490}{\km} and an along-track separation of \SI{220(50)}{\kilo\meter} \citep{Wahr2004,Kornfeld2019}. 

The \gls{GFO} twin satellites host the \glsunset{KBR}\glsentrylong{KBR} (\glsentryshort{KBR}; or \glsunset{MWI}\glsentrylong{MWI}, \gls{MWI}) and the \gls{LRI} for precisely measuring the inter-satellite distance variations \citep{Kornfeld2019}. The conventional \gls{KBR}, which is the primary ranging instrument, can resolve the inter-satellite distance variations with a noise level of about \SI{1}{\micro\meter\per\rtHz} at Fourier frequencies of \SI{1}{\hertz} \citep{Kornfeld2019}. The \gls{LRI}, a technology demonstrator, uses heterodyne near-infrared laser interferometry for the ranging measurement \citep{Sheard2012}. It has been and is still performing well after more than four years in orbit, with a noise level of \SI{200}{\pico\meter\per\rtHz} at Fourier frequencies of \SI{1}{\hertz} \citep{Abich2019}, 5000 times more precise than the \gls{KBR}.

Data processing of the inter-satellite range rate observations from \gls{KBR} or \gls{LRI} in addition to observations from the \gls{GPS} receivers, accelerometers, and star cameras as well as precise modeling of the ocean and solid Earth tides and other known effects yields monthly gravity maps of the Earth as the main scientific mission results \citep{Wahr2004,Tapley2004}. Comparing individual months and the long-term mean gravity reveals trends and annual hydrological signals for climate studies, such as accelerated ice sheet melting, groundwater storage depletion, closure of the sea-level rise budget, and more \citep{Tapley2019}. The successful commissioning of the \gls{LRI} instrument was an essential step towards the \gls{LISA} mission, which will use comparable inter-satellite laser ranging technology between three spacecraft in deep space for the detection of gravitational waves \citep{AmaroSeoane2017}.

In this paper, we investigate the ranging data of the \gls{LRI} for so-called \glspl{SEU}, which are short-lived disturbances in the phase measurement due to the interaction of charged particles or cosmic radiation with the onboard electronics. \Cref{sec::spaceEnvironment} discusses the space environment in the polar low-earth orbit and introduces different classifications of radiation effects on electronics. The \gls{LRI} architecture is explained in \cref{sec::LRIarchitecture} with special attention on the \gls{LRP}, in which the \glspl{SEU} occur. We simulate the digital filtering chain within the \gls{LRP} in \cref{sec::simulation} and create templates, which are then used to detect actual \glspl{SEU} in the measured phase data in \cref{sec::inflightData}. The identified \glspl{SEU} are discussed in \cref{sec::discussion}, and the results are summarized and concluded in \cref{sec::conclusion}.

\section{Space Environment}
\label{sec::spaceEnvironment}
The space radiation environment affects the electronics aboard spacecraft. Therefore space electronics are usually shielded or hardened against this radiation \citep{Stassi1988}. The space environment encountered by the spacecraft is influenced by Earth's magnetic field and sources from outer space. The radiation effects from the sun are characterized by its 11-year cycle, during which the sun emits a stream of particles with varying flux called the solar wind. It consists of electrons, protons, and heavy ions \citep{Nwankwo20}. Galactic cosmic rays are another source of particle flux composed of high-energy protons. They originate outside the solar system, from the depths of our galaxy \citep{Blasi2013}. The Earth's magnetic field traps these charged particles, and they follow the magnetic field lines \citep{VanAllen1959}. Depending on the species of particles, they populate different regions of the magnetic field, like the Van Allen radiation belts \citep{Bosser2017}. It is a system of two concentric belts ranging from approximately $\SI{1000}{\kilo\meter}$ to over $\SI{60000}{\kilo\meter}$ in altitude \citep{Metrailler2019}.

The probability for radiation-related incidents in space electronics is related to spatial variations of Earth's magnetic field. Over the past years, in-situ measurements were performed by several space missions and combined in the so-called CHAOS model (named after the space missions CHAMP, {\O}rsted, and SAC-C, \cite{Olsen2006}). The currently available version 7 of the CHAOS model also includes the SWARM mission results and ground data \citep{Finlay2020}. 
The region over the southern Atlantic exhibits a low magnetic field intensity at the altitude of a \gls{LEO}, which is commonly called the \gls{SAA}. Here, the inner Van Allen belt approaches Earth's surface. Like GRACE-FO, satellites in a \gls{LEO} orbit usually fly below the belt but may pass through the \gls{SAA}. It is known for its high radiation levels and is the site of frequent radiation-related events on satellite electronics. One such effect are \glspl{SEU}, occurring within the \gls{SAA} region in roughly 50\% of the total cases \citep{Zhang2021}.

When a single charged particle interacts with an electronic component like a transistor, it leaves a trail of electron-hole pairs within the semiconductor that generate a current pulse \citep{Todd2015}. This interaction either causes a hard error or a soft error: Hard errors cause severe malfunction up to defect of the device, while soft errors are temporary and non-destructive. Hence, \glspl{SEU} are soft errors. They may influence the value of the bit stored by a memory cell \citep{Todd2015}. This bitflip prevails until a new bit value is passed into the memory cell. On the other hand, a \gls{SEL} is a hard error that short circuits the electronics and can be disastrous \citep{Rivetta2001}. The \gls{GRACE} satellite, the predecessor to \gls{GFO}, experienced failure of one of the Instrument Control Units onboard one of its spacecraft in 2002, which is possibly deemed as the result of a \gls{SEL} \citep{Pritchard2002}.

\section{LRI Architecture}
\label{sec::LRIarchitecture}
The \gls{LRI} is a single instrument distributed on two equally equipped spacecraft, called \GFC and \GFD, and it measures the biased range between the spacecraft. It is operated in an active-transponder configuration \citep{Sheard2012}: One of the two units (the reference unit) sends out a laser beam with approx. \SI{25}{\milli\watt} optical power, which is stabilized to a reference cavity using the Pound-Drever-Hall technique \citep{Drever1983,Thompson2011}. The frequency of the emitted light field appears Doppler shifted by a frequency $f_\mathrm{D} < \SI{3}{\mega\hertz}$ due to the relative motion of the two spacecraft when it is sensed on the distant transponder spacecraft \citep{Sheard2012}. On the transponder unit, the incoming beam has only pico- to nanowatts of optical power due to the divergence of Gaussian beams and a small aperture at reception. The transponder laser is controlled by a feedback loop such that the incoming beam is reproduced with a well-defined phase relation but amplified in power before being sent back to the reference spacecraft. The transponder unit also intentionally introduces a frequency offset of $f_\mathrm{off}=\SI{10}{\mega\hertz}$. A second Doppler shift on the way back is sensed on the reference spacecraft. Ultimately, the interference between the local oscillator and round-trip beams is measured on the reference side and reads $f_\mathrm{R} = 2 f_\mathrm{D} + f_\mathrm{off}$ in terms of the beat frequency. Since the frequency offset $f_\mathrm{off}$ is known, range and gravity information in the form of Doppler shifts $f_\mathrm{D}$ can be extracted from the measured frequency $f_\mathrm{R}$. The \gls{LRI} on the transponder spacecraft, in principle, measures zero phase variations except for a well-defined phase ramp, due to the afore mentioned feedback loop implementing the frequency offset. %

\begin{figure*}[htb]
    \centering
    \includegraphics[width=.9\linewidth]{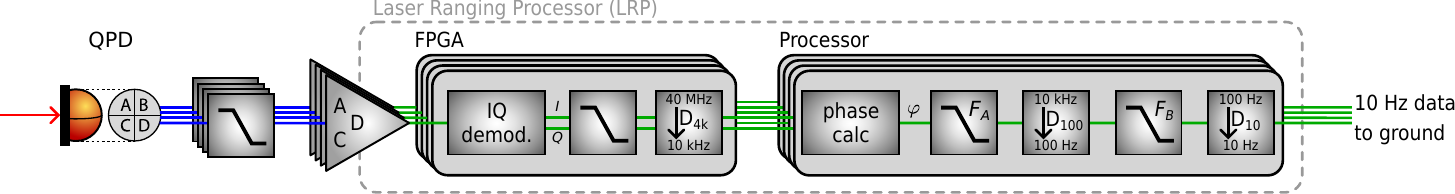}
    \caption{Phasemeter processing chain. The optical signal is converted to a voltage by the \gls{QPD} and its electronics. The voltage is filtered and digitized at a rate of approximately \SI{40}{\mega\hertz} before the \gls{FPGA}, which demodulates the signal. The phase is extracted in the processor, where further filtering and decimation takes place as well. The low-pass \gls{FIR} filters in the processor are of length $l_A$ and $l_B$, respectively, and the subsequent decimations are by a factor of 100 and 10. All shown elements are implemented independently for the four phase channels on each spacecraft. Red lines denote optical signals, blue is analog electronic and green are digital signals.}
    \label{fig::processing_chain}
\end{figure*}

Within the \gls{LRI}, the main computing engine is called the \gls{LRP}, which was built by \gls{JPL} \citep{Bachman2017}. It hosts the phase readout electronics alongside control loops for the laser, cavity, steering mirror and more. In this article, we focus on the data acquisition and processing chain, which we assume to function as depicted in \cref{fig::processing_chain}. The phase of the interfering light on both spacecraft is sensed by a \gls{QPD} allowing to retrieve ranging and beam tilt information \citep{Sheard2012} from the four phase channels per spacecraft. The photocurrents are converted into voltages within the optical bench electronics and digitized at a rate of approximately \SI{40}{\mega\hertz} before the phase information is extracted using an all-digital phase-locked loop within the \gls{LRP}, see e.\,g. \citep{Ware2006,Wand2006}. The nominal clock rates of the digitization are \SI{38.656000}{\mega\hertz} for \GFC and \SI{38.656792}{\mega\hertz} for \GFD. 
The phase extraction is divided between an \gls{FPGA}, where an IQ-demodulation, filtering and decimation to \SI{9.664}{\kilo\hertz} is performed (\SI{9.664198}{\kilo\hertz} on \GFD), and a processor, which extracts the ranging phase $\varphi = \arctan(I/Q)$, which is further decimated in a 2-step low-pass-filtering and decimation chain. The whole phase extraction and decimation chain runs individually on each of the four phase channels on both spacecraft.
The decimation in the processor comprises two \gls{FIR} filters (A and B of length $l_A$ and $l_B$) and two decimators by a factor of 100 and 10, respectively, to derive the final data rate of \SI{9.664}{\hertz} (\SI{9.664198}{\hertz} on \GFD), at which the phase data is transmitted to ground. Filtering before decimation is needed to prevent aliasing \citep{Ware2006} of higher frequencies into the measurement band of \SI{2}{\milli\hertz} to \SI{0.1}{\hertz} \citep{Dahl2017}. The two filters A and B each constitute some hundred registers (labelled $m$) and corresponding filter coefficients \citep{Ware2006} with $l_A>l_B$. The filter coefficients $c_{A/B}$ contain the impulse response of such a filter. The phase delay of all three filters adds up, giving a combined filter delay of $\SI{28802038}{clock\,ticks} \approx \SI{0.745}{\second}$ \citep{Level1UserHandbook}.
After discussions with the \gls{JPL}, the manufacturer of the \gls{LRP}, we identify the two \gls{FIR} filters A and B in the processor as the most probable source for radiation-induced \glspl{SEU} because the \gls{FPGA} can be expected to be better hardened against radiation than the memory of the processor and currently available space-qualified \glspl{FPGA} even feature error detection and correction implemented in the hardware, see e.\,g., the RTG4 FPGA Series \citep{microchip_website}.

In the following, we will use approximate values for the frequencies (e.\,g. \SI{40}{\mega\hertz} instead of \SI{38.656}{\mega\hertz}) in the text and sketches for brevity, while the simulations and data analysis uses the exact values.

\section{Simulation of Events}
\label{sec::simulation}
\begin{figure*}
    \centering
    \includegraphics[width=0.9\linewidth]{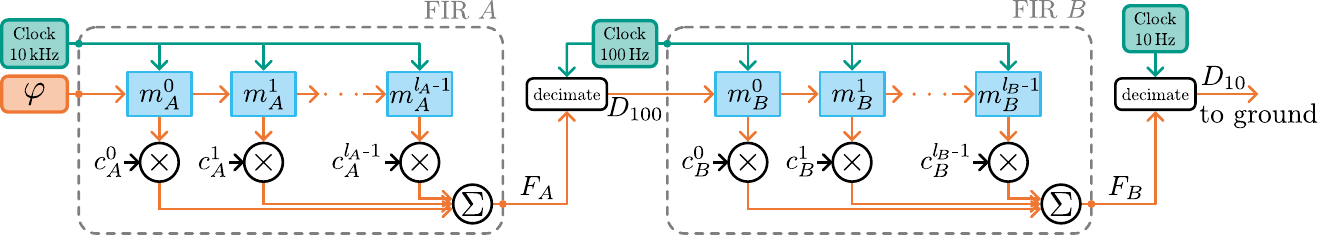}
    \caption{Block Diagram of the two FIR filter stages as implemented for the simulation. Green denotes clock signals, orange denotes the phase data and blue denotes memory cells. The FIR filter coefficients ($c_{A/B}^i$) are multiplied with the data points in the registers and the filtered result is the sum over all multiplications.}
    \label{fig::FIR_blockdiagram}
\end{figure*}

In a time-domain simulation, the output of the \gls{FIR} filtering chain was computed. 
A block diagram of the simulation is shown in \cref{fig::FIR_blockdiagram}. The filter response at a single time step is given by the sum over all the products of the register values $m_i$, containing the data $\varphi$, and their corresponding filter coefficients $c_{A/B}^i$. For the next time step, the registers values are shifted one sample to the right, and the register $m_0$ receives a new value from the input phase data.

We simulate the effect of \gls{SEU}-induced bitflips with a trivial filter input being $\varphi\equiv 0$, i.\,e., without any ranging signal, in order to obtain just the disturbance from a bitflip, and we assume that this disturbance adds to the regularly filtered signal due to linearity of \gls{FIR} filters. Hence, upon a bitflip, we set the $m$-th register from 0 to 1 during execution of the simulation.
If the \gls{SEU} occurs in filter A, it will then propagate through the subsequent filter and decimation stages. Manipulation of the $0^\mathrm{th}$ register in filter A is equivalent to setting one sample of the input phase $\varphi$ to one. However, manipulation of higher registers can not be replaced by equivalent input data $\varphi$. All intermediate data streams are computed, where \F{A} denotes the output of the first filter, which is then decimated by a factor of 100 (denoted \D{100}). The second filter output is \F{B}, and its decimated outcome at a \SI{10}{\hertz} data rate is called \D{10}. 

We identified the defining parameters of a bitflip to be
\begin{itemize}
    \item The affected filter (A or B).
    \item The occurrence time of the \gls{SEU}, expressed as a sample number or tick $k_\mathrm{A/B}$ at the filter's clock rate. Due to the fixed decimation rates from \F{A} or \F{B} to the \SI{10}{\hertz} output data rate ($1000=100\cdot 10$ or 10, respectively), the output of a varying $k$ repeats. Thus, if the \gls{SEU} occurs in filter A we use $k_\mathrm{A}\in[0,1000)\subseteq\mathbb{N}_0$ and for an \gls{SEU} in filter B we use $k_\mathrm{B}\in[0,10)\subseteq\mathbb{N}_0$. Now, $k$ can be regarded as the sub-sample time in between of two data samples of the \SI{10}{\hertz} output data.
    \item The affected register number $m_\mathrm{A}\!\in[0,l_\mathrm{A})\subseteq\mathbb{N}_0$ or $m_\mathrm{B}\!\in[0,l_\mathrm{B})\subseteq\mathbb{N}_0$ of the filter. We usually provide this number in \% of the full filter length $l_\mathrm{A/B}$.
    \item The bit number $b\in[0,64)\subseteq\mathbb{N}_0$ that flipped of the presumed 64-bit register (i.\,e. the $2^b$ magnitude of the flipped bit). For simulation, $b=0$ is usually used, since this parameter is a linear scale factor that can easily be estimated through a least squares algorithm. %
\end{itemize}
We simulate the bitflips with $\varphi=0$ as initial condition and the bit flipping from zero to one. However, one could also initialize $\varphi=1$ and flip from one to zero. This results in the same shapes of the output data but with inverted sign.
\begin{figure}
    \centering
    \includegraphics[width=\linewidth]{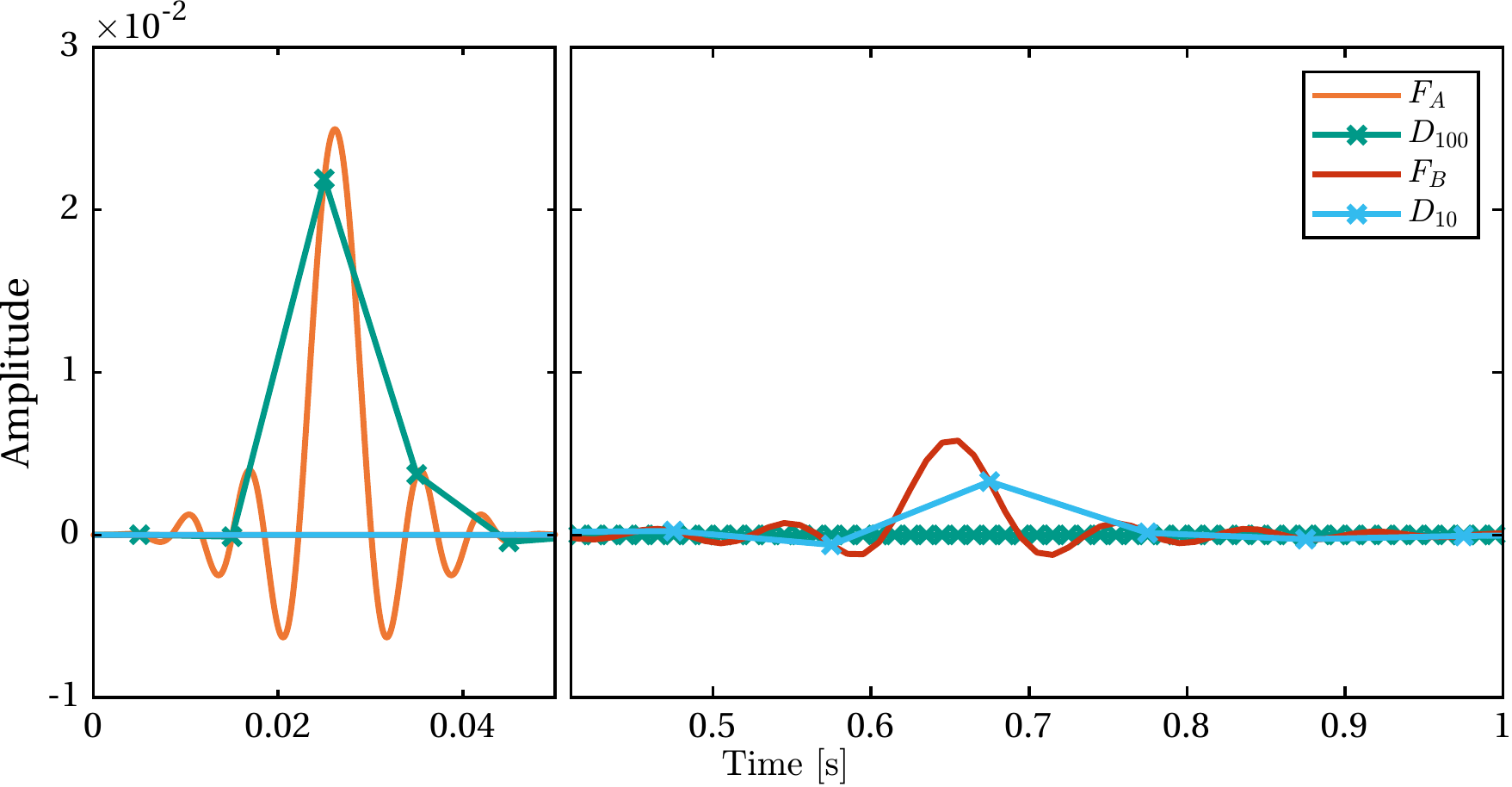}
    \caption{Simulated data throughout the filtering chain for an \gls{SEU} in the first \gls{FIR} filter with injection sample and register number $k=m=0$ and a magnitude $a=1$. The input data $\varphi$ is zero and thus not shown. The output of the first filter \F{A} (orange) is sampled at \SI{10}{\kilo\hertz}, the first decimation \D{100} (green) and the output of the second filter \F{B} (red) at \SI{100}{\hertz} and the final output \D{10} (cyan) is sampled at \SI{10}{\hertz}. Both time-axes are in units of seconds, but note the different scale. These examples show artificial filter coefficients, as the exact coefficients employed in-flight can unfortunately not be disclosed here.}
    \label{fig::simulated_data_k1_m1}
\end{figure}
\Cref{fig::simulated_data_k1_m1} shows an exemplary simulation result for an \gls{SEU} in the first register ($m=0$) of the first filter (A) at time $k_A=0$. Orange and green are the intermediate data streams after the first filter, red is the second filter's output, and cyan is the final \SI{10}{\hertz} output data. A larger injection sample, $k_A>0$, would cause a delay of \F{A} and thus a slightly different shape and amplitude of the subsequent data streams due to the different sampling of \F{B}. 

\begin{figure}
    \centering
    \includegraphics[width=\linewidth]{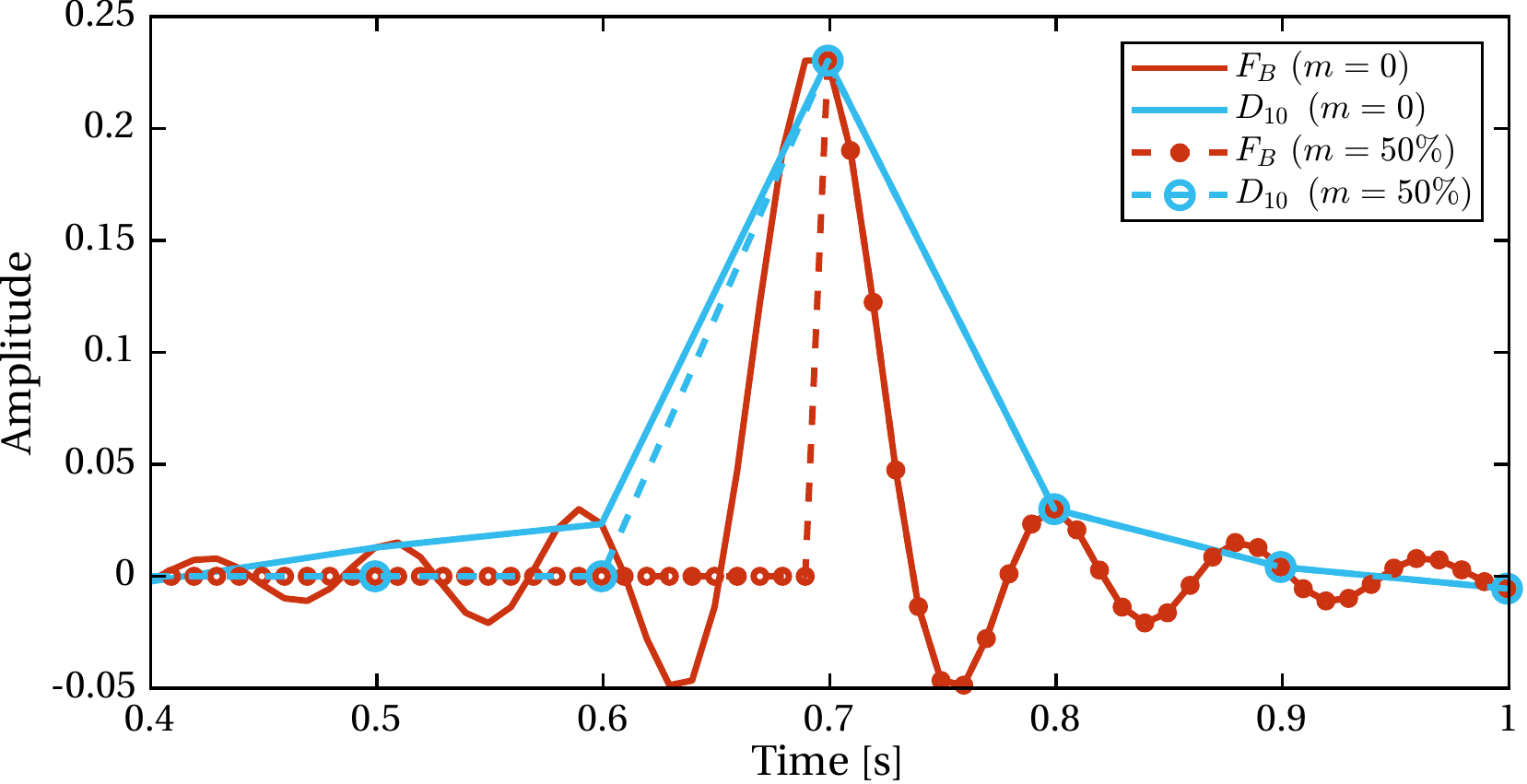}
    \caption{Simulated data showing the effect of an \gls{SEU} in a higher register number for the second filter. The solid lines depict the response of an impulse travelling through the full filter (i.\,e., for $m=0$), while the dashed lines show the response for an \gls{SEU} that affects the register $m=50\%$ in the middle of the filter. Color coding as in \cref{fig::simulated_data_k1_m1}. Note, that the magnitude here is larger than in \cref{fig::simulated_data_k1_m1}, because this \gls{SEU} was simulated in the second instead of the first \gls{FIR} filter. These examples show artificial filter coefficients, as the exact coefficients employed in-flight can unfortunately not be disclosed here.}
    \label{fig::simulated_data_k1_m70}
\end{figure}

When a low register number $m$ is affected by the bitflip it implies that almost the complete filter impulse response is visible in the immediate output, as shown by the solid lines of \cref{fig::simulated_data_k1_m70}, where the red line depicts the immediate output of filter B at \SI{100}{\hertz} and blue denotes the decimated data at \SI{10}{\hertz}. A higher register number $m$ yields cropped filter responses in the immediate output, as shown by the dashed lines in \cref{fig::simulated_data_k1_m70}. For a high register number $m$ in filter A, the \SI{10}{\hertz} output data would not appear cropped since the cropped and decimated filter output \D{100} is filtered once more in \F{B}, which ultimately dominates the shape of the output \D{10}. 

For a fixed register number $m$, the output \D{10} can have very different shapes, depending on the time or sample $k$ at which the \gls{SEU} was induced in the data. There are 1000 unique patterns in the \SI{10}{\hertz} output data stream for an \gls{SEU} in the first filter A and ten patterns for the second filter B, according to the sampling rate decimation factors. The ten patterns of filter B are approximately a subset of the 1000 patterns of filter A since the output $D_{100}$ of filter A is approximately only a single peak which is then fed into filter B. 

Two \glspl{LUT} for events either in filter A or B were created from the simulations, where the injection sample number $k$ and the register number $m$ at which the \gls{SEU} was injected into the filter were varied over the parameter space. The resulting output data after the second decimation, i.\,e., at \SI{10}{\hertz}, is stored in the \glspl{LUT}. The full 3D-\glspl{LUT} have the dimensions $1000 \times l_A \times 15$ and $10 \times l_B \times 15$, respectively, where the first dimension represents the injection sample number $k$, the second dimension is the register number $m$ and the third dimension is the total number of data points of the complete filter response at \SI{10}{\hertz} in \D{10}. The individual rows of the \glspl{LUT} are denoted as \LUT{A/B}{k}{m}. For better readability, we will omit the subscript A/B in the following, where we usually mean that all the equations are evaluated independently for both \glspl{LUT}.

Since the true \gls{LRP}-internal filter coefficients are only available project-internally, we use exemplary \gls{FIR} filters to show the principle in \cref{fig::simulated_data_k1_m1,fig::simulated_data_k1_m70}. The following analysis of actual flight-data however uses the true in-flight \gls{LRP} filter coefficients.

\section{Detection of SEUs in LRI Phase Data}
\label{sec::inflightData}
The \gls{SEU} detection algorithm is part of a larger framework developed at the Albert-Einstein-Institute in Hanover to automatically process and analyze \gls{LRI} data in near-real-time \citep{Misfeldt2019}. It features an outlier-detection, originally developed to remove thruster-induced phase jumps \citep{Abich2019} but was now extended to identify \glspl{SEU}. The overall process is two-fold: first, all phase disturbance events are detected and categorized. All events where the first derivative of the measured phase (or the phase rate) exhibits steps larger than \SI{\pm30}{\milli\hertz}, or where the \gls{DWS} combination shows outliers larger than \SI{2e-4}{cycles\per\second}, are marked as potential phase disturbance events. Subsequently, modeling and subtraction are performed. The criterion for deciding whether a phase disturbance is an \gls{SEU} and not a true phase jump due to optical or mechanical disturbances is that an \gls{SEU} occurs in a single channel only since the filtering and decimation of the four channels are performed separately. In contrast, a phase jump affects all four channels, and further, an \gls{SEU} produces a short-lived peak (after propagating through the filter, the disturbance vanishes), while a phase jump causes a persistent step in the ranging data (caused by a non-zero integral of fast laser frequency variations, cf. \cite{Misfeldt2019}). A short segment of $N \leq 30$ samples of the affected channel is extracted from the measured phase data once an \gls{SEU} candidate is identified. The mean over the three unaffected channels is subtracted from the affected channel to remove the common (ranging) signal and extract a clean signature of the \gls{SEU}. We call this extracted bitflip signal $\varphi(t_i)$ or $\varphi_i$, where $t_i$ are the discrete-time samples and $i$ is the sample number. Exemplarily, if an \gls{SEU} occurs in channel C, then $\varphi(t_i) = \varphi_C(t_i) - \left(\varphi_A(t_i)+\varphi_B(t_i)+\varphi_D(t_i)\right)/3$. This expression additionally suppresses common-mode noises like laser frequency noise (on the reference side). The measurement noise of the \gls{LRI} is discussed in more detail in \citet{Mueller2022}.

We introduce our model for the \gls{SEU} phase
\begin{equation}
    \eta_i^{k,m}(a) = \eta^{k,m}(a, t_i) = a\cdot\LUT{}{k}{m}(t_i)\ ,
\end{equation}
which essentially is an \gls{LUT} entry scaled by an amplitude $a$, and the residuals
\begin{align}
    r_i^{k,m}(\boldsymbol\vartheta) &= r\left((a, c_2, c_1, c_0)^\transpose, t_i\right) \notag \\
    &=  \varphi_i - \eta_i^{k,m}(a)  - c_2\cdot t_i^2 - c_1\cdot t_i - c_0\ ,
\end{align}
where we further subtract a second order polynomial, which may still be present in the data $\varphi$ due to insufficient (ranging) signal removal or similar effects. This equation defines the regression coefficients $\boldsymbol\vartheta = (a,c_2,c_1,c_0)^\transpose$.

To assess which of the $k\times m$ models in the \glspl{LUT} matches the data best, we employ the framework of maximum likelihood estimation. First, we compute the likelihood of $\boldsymbol\vartheta$ given the measured data $\varphi$ as \citep{Koch1999}
\begin{equation}
    \pdfunc[\,k,m]{\varphi}{\boldsymbol\vartheta} = \frac{1}{\sqrt{\left|2\pi\Sigma\right|}}\cdot \exp\left( -\frac{1}{2}r^{k,m}(\boldsymbol\vartheta)^\transpose\ \Sigma^{-1}\ r^{k,m}(\boldsymbol\vartheta) \right)\ .
\end{equation}
The covariance matrix $\Sigma$ will be discussed later. The best fitting model $\eta^{k,m}(a)$ can be identified by the maximum value of the likelihood function $\mathcal{L}$ over the parameter space or equivalently by the minimum of its negative logarithm 
\begin{align}
    \loglikeli[\,k,m]{\varphi}{\boldsymbol\vartheta} &= - \ln \pdfunc[\,k,m]{\varphi}{\boldsymbol\vartheta} \\
        &= \frac{1}{2}\ln\left(\left|2\pi\Sigma\right|\right) + \frac{1}{2}r^{k,m}(\boldsymbol\vartheta)^\transpose\cdot\Sigma^{-1} \cdot r^{k,m}(\boldsymbol\vartheta)\ . 
    \label{eq::log-likelihood}
\end{align}
The parameter space is discrete for the parameters $k$ and $m$ and continuous for $\boldsymbol\vartheta$. Hence we minimize the negative log-likelihood $\ell^{k,m}$ for all $k,\,m$ through a generalized least squares, i.\,e., by estimating
\begin{equation}
    \hat{\boldsymbol\vartheta} = \underset{\boldsymbol\vartheta}{\mathrm{argmin}}\ r^{k,m}(\boldsymbol\vartheta)^\transpose\cdot\Sigma^{-1}\cdot r^{k,m}(\boldsymbol\vartheta)\ . \label{eq::GLSQ}
\end{equation}
Ultimately, the best estimate for the \gls{SEU} model is determined by finding the minimum of $\loglikeli[\,k,m]{\varphi}{\hat{\boldsymbol\vartheta}}$ in the two-dimensional $k\times m$-sized grid.

\begin{figure}
    \centering
    \includegraphics[width=\linewidth]{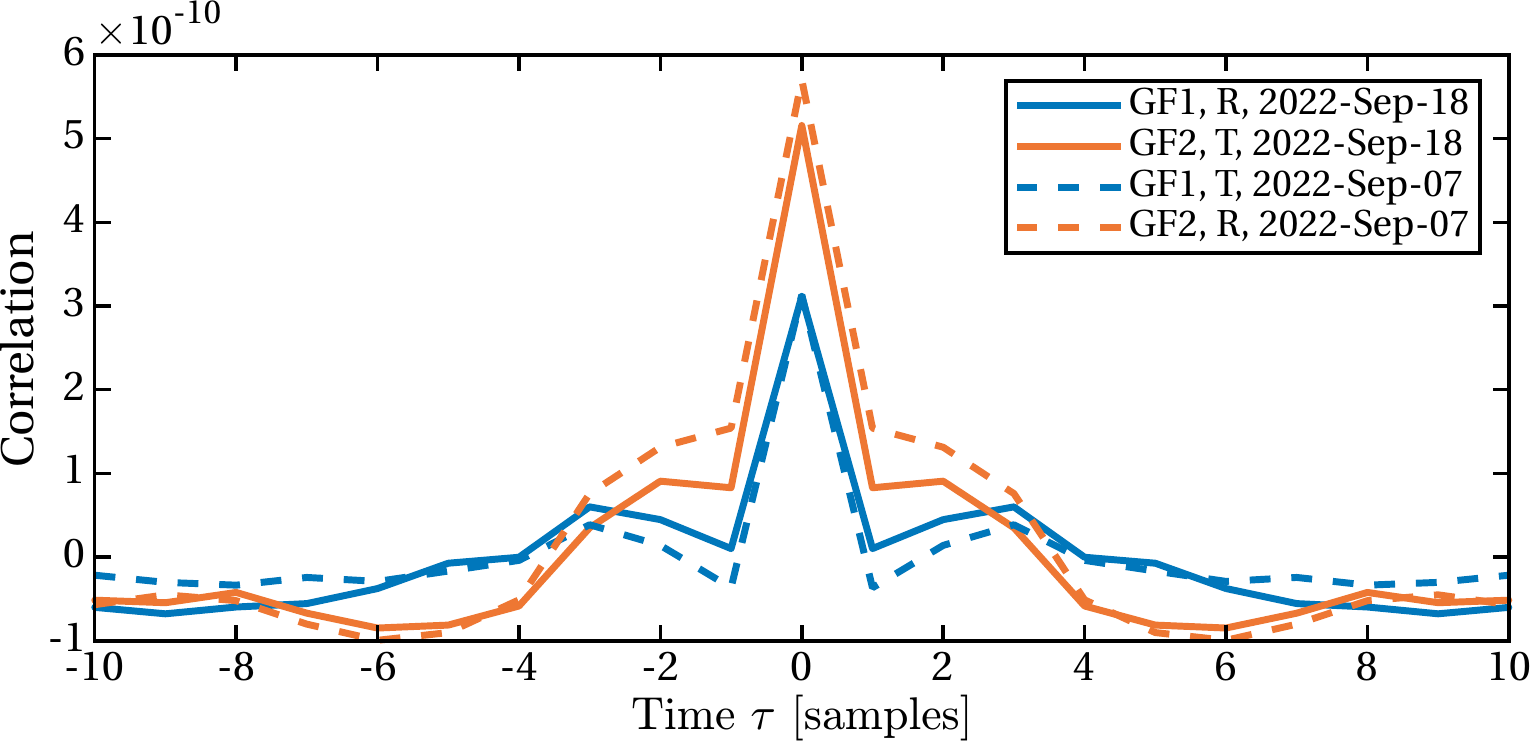}
    \caption{Exemplary autocorrelation for a single channel phase combination $\varphi_A-(\varphi_B+\varphi_C+\varphi_D)/3$ of \GFC and \GFD on two different days with different roles.}
    \label{fig::autocorrelation}
\end{figure}

The covariance matrix $\Sigma$, which is needed to compute the generalized least squares (cf. \cref{eq::GLSQ}), is derived from the expectation value $\mathrm{E}$ of the measurement noise $n$ as
\begin{equation}
    \Sigma_{ij} = \mathrm{E}[n_i \cdot n_j] = R_n(t_i-t_j)\ .
    \label{eq::covariance_matrix}
\end{equation}
Here, the expectation value $\mathrm{E}$ can be computed through the unbiased correlation function of the (real-valued) data $n$ of length $N$ as
\begin{equation}
    R_n(\tau) = \left\{
        \begin{array}{ll}
            \displaystyle \frac{1}{N-\tau}  \sum_{i=0}^{N-\tau-1}n_{i+\tau}n_i\,, & \tau \geq 0 \\[1.2em]
            R_n(-\tau)\,, & \tau < 0\ .
        \end{array}
    \right. 
\end{equation}
The correlation function is obtained from the autocorrelation of actual phase data in absence of an \gls{SEU} event. Shown in  \cref{fig::autocorrelation} is the mean over \num{20000} autocorrelations of consecutive data segments with 30 samples length for the two spacecraft in both roles. A trend was removed from the phase data before computing each autocorrelation.
The function differs a bit in their shape between \GFC and \GFD, and the magnitude varies insignificantly between different days in different roles (reference or transponder). The values of the solid lines are used as the correlation function $R_n(\tau)$ to form the covariance matrix $\Sigma$ from the measurement noise $n$.

From the fitted amplitudes $a$ of the \gls{LUT} rows, we directly obtain the amplitude and sign of the \gls{SEU} as it occurred before the filtering. We can further compute the bit number $b$ of the affected bit by 
\begin{equation}
    b = \log_2\left(10\cdot 2^{24}\cdot a\right)\ , \label{eq::bit_significance}
\end{equation}
where $1/(10\cdot2^{24})$ is the least significant bit in units of phase cycles in the \gls{LRI} phase measurement \citep{Level1UserHandbook}. Ideally, $b$ yields an integer number.

The above computation is done individually for all templates in the two filter's \glspl{LUT} (\LUT{A}{k}{m} and \LUT{B}{k}{m}). We compare the two minimal values of the log-likelihood over the \glspl{LUT} to identify the most likely filter (A or B) when the \gls{SEU} occurred.

\section{Discussion}
\label{sec::discussion}
\begin{figure}[bt]
    \centering
    \includegraphics[width=\linewidth]{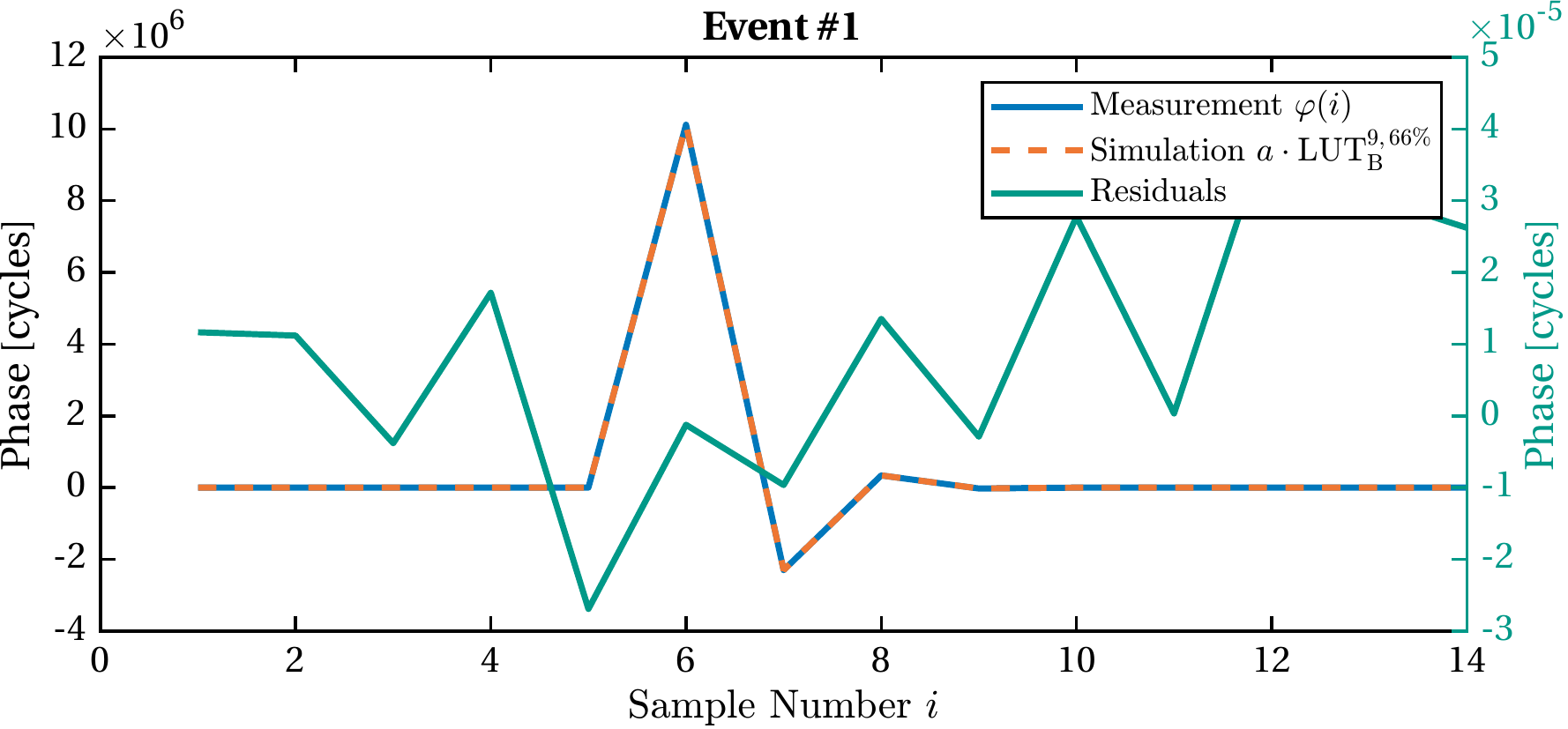}
    \caption{Event \#1: Example of a good \gls{SEU} fitting result. The blue trace shows the isolated segment from the phase data of channel D on \GFD on 2018-July-09 around 18:25 UTC. An \gls{SEU} in bit $b=60$ of the register $67\%\cdot l_B$ of the filter B (dashed orange) was subtracted, which yields the green residuals (scale according to right axis). The noise after subtraction is in the order of \SI{e-5}{cycles}.}
    \label{fig::usualSeuPlot}
\end{figure}

\begin{table*}[b]
    \small
    \centering
    \caption{SEU parameters as detected in the LRI phase data. Ch denotes the affected phase channel; $k$ is the sub-sample timing of the event; $m$ is the affected register (normalized by the total number of registers $l_\mathrm{A/B}$); $b$ is the bit that flipped; Dir denotes the direction of the bitflip ($0\rightarrow1$ ($\uparrow$) or $1\rightarrow0$ ($\downarrow$). A 95\% confidence interval (CI) for the bit number, derived from the formal errors of the least squares fit and assuming a Gaussian distribution, is computed as \num{1.96} times the standard deviation of the noise. The colored bit position cells are encoded as follows: \colorbox{pale_green}{Green:} Fractional number can be well explained with two bitflips at the same time. \colorbox{pale_yellow}{Yellow:} Fractional number can possibly be explained with more than two bitflips at the same time. \colorbox{pale_gray}{Gray:} High residuals observed (see main text, \cref{sec::otherEvents}). The horizontal lines separate different years.}
    \label{tab::SEUfitResults}
    
    \begin{tabular}{
        cccccc %
        S[table-format=03] %
        S[table-format=03] %
        >{$}c<{$} %
        S[table-format=2]
            @{}>{\columncolor{white}[0pt][\tabcolsep]} %
        S[table-format=-1.2e2,retain-zero-exponent=true,scientific-notation=true,explicit-sign=+] %
        S[table-format=-1.2e2,retain-zero-exponent=true,scientific-notation=true] %
        S[table-format=1.2e2,retain-zero-exponent=true,scientific-notation=true]
    }
        \hline \\[-0.85em]
        \# & Event Time & SC & Role & Ch & FIR & $k$ & $m$ & \mathrm{Dir} & \multicolumn{2}{c}{Bit No. $b$} & {95\% CI} & {Residuals} \\
         & [UTC] & & & & & & [\%] & & \multicolumn{2}{c}{int+frac} & {Bit No. $b$} & {[cycles rms]} \\[0.15em]
        \hline \\[-0.85em]
        1 & 09-Jul-2018 18:25:01 & \GFC & T & D & B & 9 & 66 & \uparrow &  60 &  1.25e-12 & \pm 4.29e-12 &  1.33e-05 \\[0em] \\[-0.85em]
        \hline \\[-0.85em]
        2 & 20-Jan-2019 09:31:55 & \GFC & R & D & A & 296 & 31 & \uparrow & \cellcolor{pale_yellow} 35 & \cellcolor{pale_yellow} -1.97e-02 & \pm 1.49e-04 &  9.09e-06 \\
        3 & 01-Apr-2019 16:36:27 & \GFD & T & B & A & 352 & 0 & \downarrow &  46 &  -2.73e-04 & \pm 6.20e-04 & \cellcolor{pale_gray}  7.42e-02 \\
        4 & 18-Apr-2019 23:38:38 & \GFD & T & C & A & 165 & 37 & \uparrow &  63 &  8.81e-13 & \pm 2.13e-12 &  2.53e-05 \\
        5 & 09-Sep-2019 00:07:28 & \GFD & T & A & B & 4 & 6 & \downarrow & \cellcolor{pale_green} 30 & \cellcolor{pale_green} 8.75e-02 & \pm 3.59e-05 &  7.36e-06 \\
        6 & 13-Nov-2019 23:31:24 & \GFC & R & D & A & 316 & 67 & \downarrow & \cellcolor{pale_yellow} 53 & \cellcolor{pale_yellow} 2.12e-01 & \pm 2.43e-08 &  1.03e-05 \\
        7 & 20-Nov-2019 03:09:38 & \GFC & R & A & A & 309 & 49 & \downarrow &  50 &  6.70e-08 & \pm 1.07e-07 &  1.11e-05 \\
        8 & 25-Nov-2019 15:30:17 & \GFC & R & C & B & 1 & 65 & \downarrow &  34 &  -8.10e-05 & \pm 3.59e-04 &  8.39e-06 \\[0em] \\[-0.85em]
        \hline \\[-0.85em]
        9 & 06-Jan-2020 06:13:57 & \GFD & T & C & B & 4 & 56 & \downarrow &  41 &  1.33e-01 & \pm 4.87e-03 & \cellcolor{pale_gray}  1.71e-01 \\
        10 & 01-Feb-2020 16:24:25 & \GFD & T & A & A & 299 & 31 & \uparrow & \cellcolor{pale_yellow} 33 & \cellcolor{pale_yellow} -1.59e-01 & \pm 4.42e-04 &  5.66e-06 \\
        11 & 22-Mar-2020 00:51:54 & \GFC & R & A & A & 797 & 51 & \uparrow & \cellcolor{pale_yellow} 35 & \cellcolor{pale_yellow} -1.51e-01 & \pm 1.46e-02 &  3.88e-05 \\
        12 & 01-May-2020 08:07:03 & \GFC & R & C & A & 646 & 0 & \downarrow & \cellcolor{pale_yellow} 32 & \cellcolor{pale_yellow} -9.28e-04 & \pm 1.35e-03 &  1.06e-05 \\
        13 & 30-May-2020 09:31:44 & \GFC & R & A & A & 394 & 32 & \downarrow &  59 &  2.92e-12 & \pm 1.18e-11 &  1.06e-05 \\
        14 & 17-Aug-2020 23:27:50 & \GFC & R & A & B & 1 & 86 & \uparrow & \cellcolor{pale_yellow} 39 & \cellcolor{pale_yellow} 2.56e-01 & \pm 2.54e-03 &  8.27e-06 \\
        15 & 07-Sep-2020 01:47:43 & \GFD & T & A & B & 6 & 0 & \uparrow &  37 &  -4.50e-02 & \pm 1.09e-06 &  2.35e-05 \\
        16 & 12-Sep-2020 02:17:38 & \GFC & R & B & A & 974 & 16 & \uparrow & \cellcolor{pale_green} 29 & \cellcolor{pale_green} -1.12e-02 & \pm 8.84e-03 &  8.89e-06 \\
        17 & 11-Dec-2020 18:06:47 & \GFD & T & D & A & 903 & 30 & \uparrow & \cellcolor{pale_green} 49 & \cellcolor{pale_green} -6.70e-09 & \pm 7.84e-09 &  9.22e-06 \\
        18 & 19-Dec-2020 04:44:29 & \GFD & T & B & A & 370 & 76 & \downarrow & \cellcolor{pale_yellow} 54 & \cellcolor{pale_yellow} 2.73e-01 & \pm 8.71e-09 &  7.46e-06 \\[0em] \\[-0.85em]
        \hline \\[-0.85em]
        19 & 09-Mar-2021 11:18:40 & \GFD & T & D & B & 3 & 35 & \downarrow &  51 &  1.23e-11 & \pm 3.96e-11 &  1.61e-05 \\
        20 & 10-Mar-2021 18:20:28 & \GFD & T & D & A & 218 & 47 & \downarrow &  39 &  8.81e-05 & \pm 8.57e-05 &  1.22e-05 \\
        21 & 12-Mar-2021 23:20:11 & \GFD & R & C & A & 434 & 24 & \uparrow & \cellcolor{pale_green} 37 & \cellcolor{pale_green} -3.24e-04 & \pm 2.90e-05 &  8.18e-06 \\
        22 & 25-Jul-2021 00:27:50 & \GFD & T & B & B & 0 & 23 & \uparrow &  51 &  8.50e-11 & \pm 6.20e-11 &  2.58e-05 \\
        23 & 30-Sep-2021 19:46:16 & \GFD & T & A & A & 57 & 0 & \uparrow &  44 &  -3.39e-08 & \pm 5.13e-07 &  2.13e-05 \\
        24 & 16-Nov-2021 16:25:26 & \GFD & T & C & A & 18 & 7 & \downarrow &  45 &  1.53e-07 & \pm 3.65e-07 &  2.23e-05 \\
        25 & 05-Dec-2021 04:06:04 & \GFC & R & C & A & 454 & 60 & \uparrow & \cellcolor{pale_green} 38 & \cellcolor{pale_green} -3.05e-02 & \pm 3.05e-04 &  9.73e-06 \\[0em] \\[-0.85em]
        \hline \\[-0.85em]
        26 & 27-Jan-2022 22:21:58 & \GFC & R & B & B & 7 & 32 & \downarrow &  51 &  -1.75e-11 & \pm 1.84e-11 &  7.06e-06 \\
        27 & 04-Jun-2022 15:21:35 & \GFC & T & D & A & 721 & 46 & \uparrow &  63 &  -3.03e-12 & \pm 3.70e-12 &  1.09e-05 \\
        28 & 07-Nov-2022 12:32:14 & \GFC & R & C & B & 9 & 66 & \downarrow & \cellcolor{pale_green} 29 & \cellcolor{pale_green} 8.51e-03 & \pm 1.20e-02 &  1.91e-05 \\
        29 & 22-Nov-2022 22:27:29 & \GFD & T & B & A & 542 & 2 & \uparrow &  63 &  3.34e-13 & \pm 1.23e-12 &  1.96e-05 \\
        \hline
    \end{tabular}
\end{table*}
Over the analyzed mission time ranging from June 2018 until the end of December 2022, in which the \gls{LRI} was in science mode for more than 75\% of the time, we identified 29 \gls{SEU} events in the \gls{LRI} phase data, whose parameters are shown in \cref{tab::SEUfitResults}. A time series of an exemplary \gls{SEU} event (\#1 in the table), the fitted model, and the corresponding residuals are shown in \cref{fig::usualSeuPlot}.

Of all events, \GFC recorded 14 events, while \GFD recorded 15 events. As the reference/transponder role can be switched, 16 were detected on the transponder unit, and 13 on the reference unit of the \gls{LRI}. The distribution over the four channels is almost equal (A: 8 events, B: 6, C: 8, D: 7). Filter A shows more events (19 vs. 10 in filter B). This is expected, since filter A has more registers $l_A > l_B$, i.\,e., a physically larger area in the electronics that can be hit by radiation. Given that the \gls{LRI} was in science mode for more than 85\% of the time in 2019--2021, approximately nine events can be expected annually. It was not in science mode for long periods in 2018 and 2022; thus, fewer events were observed. The smallest observed event occurred in the 29$^\mathrm{th}$ bit (event \#28). Thus we expect that there are actually more \gls{SEU} events at low bit numbers, but they are not detectable in the \gls{LRI} noise.

The subtraction of the \gls{SEU} signature from ranging data, in general, works well since the rms of the residuals is in the order of some \SI{e-5}{cycles} in most of the cases, which is the noise level of the phase measurement system (cf. \citet{Mueller2022}).

\begin{figure}
    \centering
    \includegraphics[width=\linewidth]{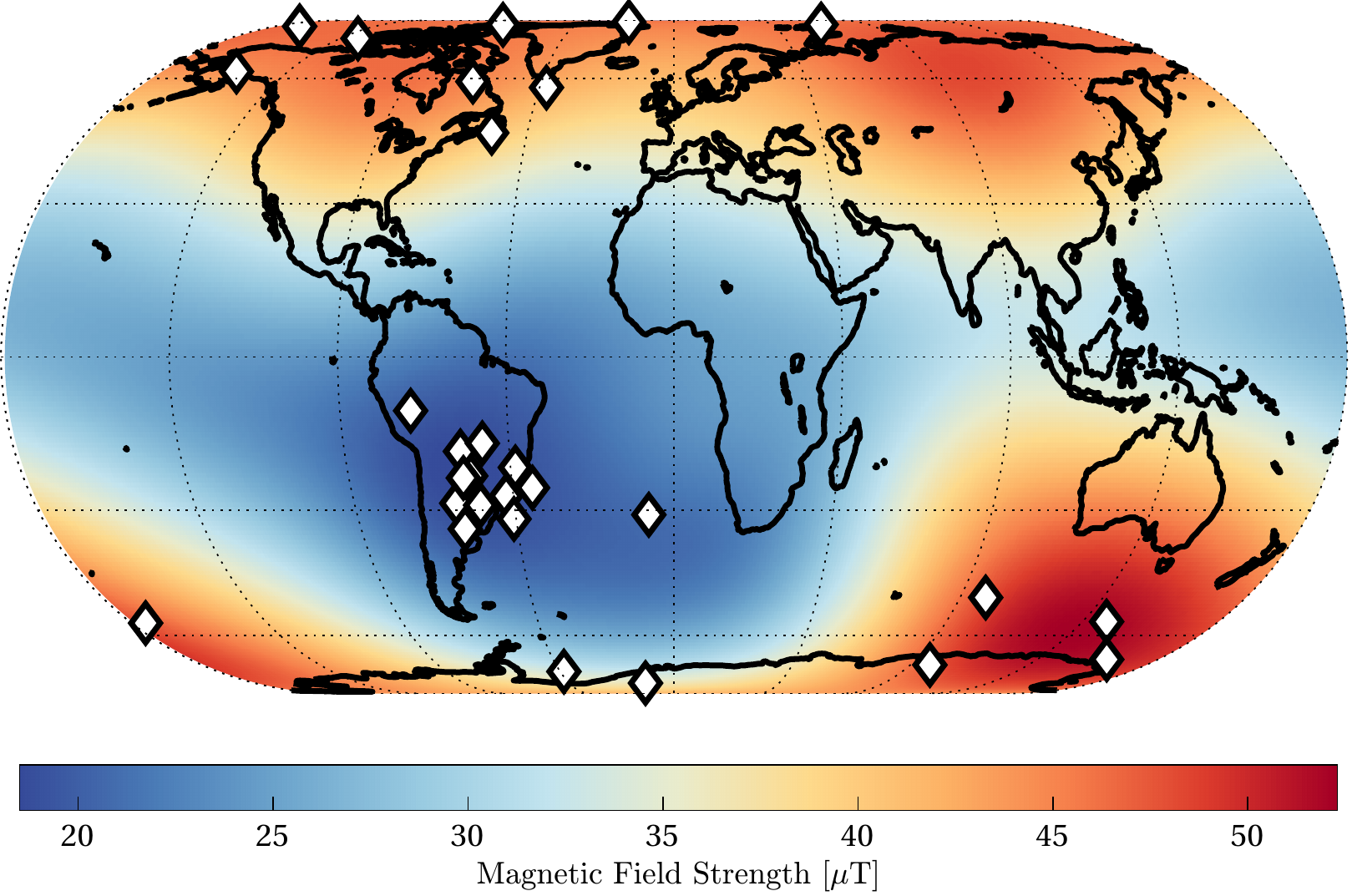}
    \caption{World map showing the location of the \gls{GFO} spacecraft at occurrence of \glspl{SEU} (diamonds). The color coding depicts the strength of the magnetic field in \si{\micro\tesla} at an altitude of \SI{490}{\kilo\meter} above Earth's surface, as derived from the CHAOS-7 model for January 2021 \citep{Finlay2020}. There is evidence for an increased number of \glspl{SEU} in the region of the South-Atlantic Anomaly.}
    \label{fig::Chaos_SAA}
\end{figure}
The distribution of the ground-track position of the spacecraft at the time of the \gls{SEU} events (shown in \cref{fig::Chaos_SAA}) reveals an expected clustering within the South-Atlantic Anomaly, where almost 50\% of the events take place. This is consistent with results from the literature \citep{Zhang2021}.

We did explicitly exclude the possibility, that an \gls{SEU} could also alter the filter coefficients. A bitflip in the coefficients would cause a different filter gain and noise suppression. However, the exact effects also strongly depend on the architecture and implementation in the \gls{LRP}.

\subsection{Non-Integer Bit Numbers}
Some events show bit number $b$, that are not integer within the 95\% confidence interval, marked with different colors in \cref{tab::SEUfitResults}. Though non-integer bit numbers seem contra-intuitive in the first place, it can be explained when considering a simultaneous bitflip in separate bits. This increases or decreases the signal's amplitude and thus the retrieved bit number $b$ (cf. \cref{eq::bit_significance}). The allowed fractional bit numbers obtained from our fit only depend on the separation in bits between the affected bits:
\begin{equation}
    \mathcal{O}_\pm(n) = \log_2(2^b \pm 2^{b-n})-\log_2(2^b)\ .
\end{equation}
The sign of the $2^{b-n}$-term denotes the direction of the lower bit at position $b\!-\!n$ with respect to the flip direction of the upper bit $b$, which is indicated in \cref{tab::SEUfitResults}.
The first 12 allowed fractional bit values are shown in \cref{tab::magic_numbers}. 
\begin{table}[tb]
    \centering
    \caption{Fractional bit number for two bitflips at the same time as a function of the separation between bit numbers. The number $n$ denotes position $b-n$ of the second bit, relative to the one at position $b$, $b>n$.}
    \label{tab::magic_numbers}
    \begin{tabular}{c>{$}c<{$}>{$}c<{$}}
        \hline
        $n$ & \mathcal{O}_+(n) & \mathcal{O}_-(n) \\
        \hline
        1 & 0.58496 & -1 \\
        2 & 0.32193 & -0.41504 \\
        3 & 0.16993 & -0.19265 \\
        4 & 0.08746 & -0.09311\\
        5 & 0.04439 & -0.04580\\
        6 & 0.02237 & -0.02272\\
        7 & 0.01123 & -0.01131\\
        8 & 0.00562 & -0.00565\\
        9 & 0.00282 & -0.00282\\
        10 & 0.00141 & -0.00141\\
        11 & 0.00070 & -0.00070 \\
        12 & 0.00035 & -0.00035 \\
        \hline
    \end{tabular}
\end{table}
Note that $\mathcal{O}_+(1)$ and $\mathcal{O}_-(2)$ are degenerate and also a flip in bit $b$ and $b\!-\!1$ in different directions can not be distinguished from a single flip in the $b\!-\!1$-th bit. Comparing the allowed fractional bit numbers from \cref{tab::magic_numbers} with the values in column ``Bit No. $b$" of \cref{tab::SEUfitResults}, several events can be explained by multiple bitflips. For instance, we observe $\mathcal{O}_+(4)$ (the 30th and 26th bit flipped in the same direction) for event \#5. All these events are marked green. The numbers of \cref{tab::magic_numbers} can not directly explain the events marked yellow. However, these fractional bit positions can be explained when considering even more than two bitflips simultaneously. The fractional bit number of event \#10 is $\num{-0.159}\approx\mathcal{O}_-(3)+\mathcal{O}_+(5)$, which denotes bitflips in the 31$^\mathrm{st}$, 28$^\mathrm{th}$ and 26$^\mathrm{th}$ bit.
Any assessment of how likely a single particle's impact may induce two bits to flip strongly depends on the exact architecture and physical arrangement of the memory cells, which is unknown to the authors.

\subsection{Other Events} \label{sec::otherEvents}
\begin{figure}
    \centering
    \includegraphics[width=\linewidth]{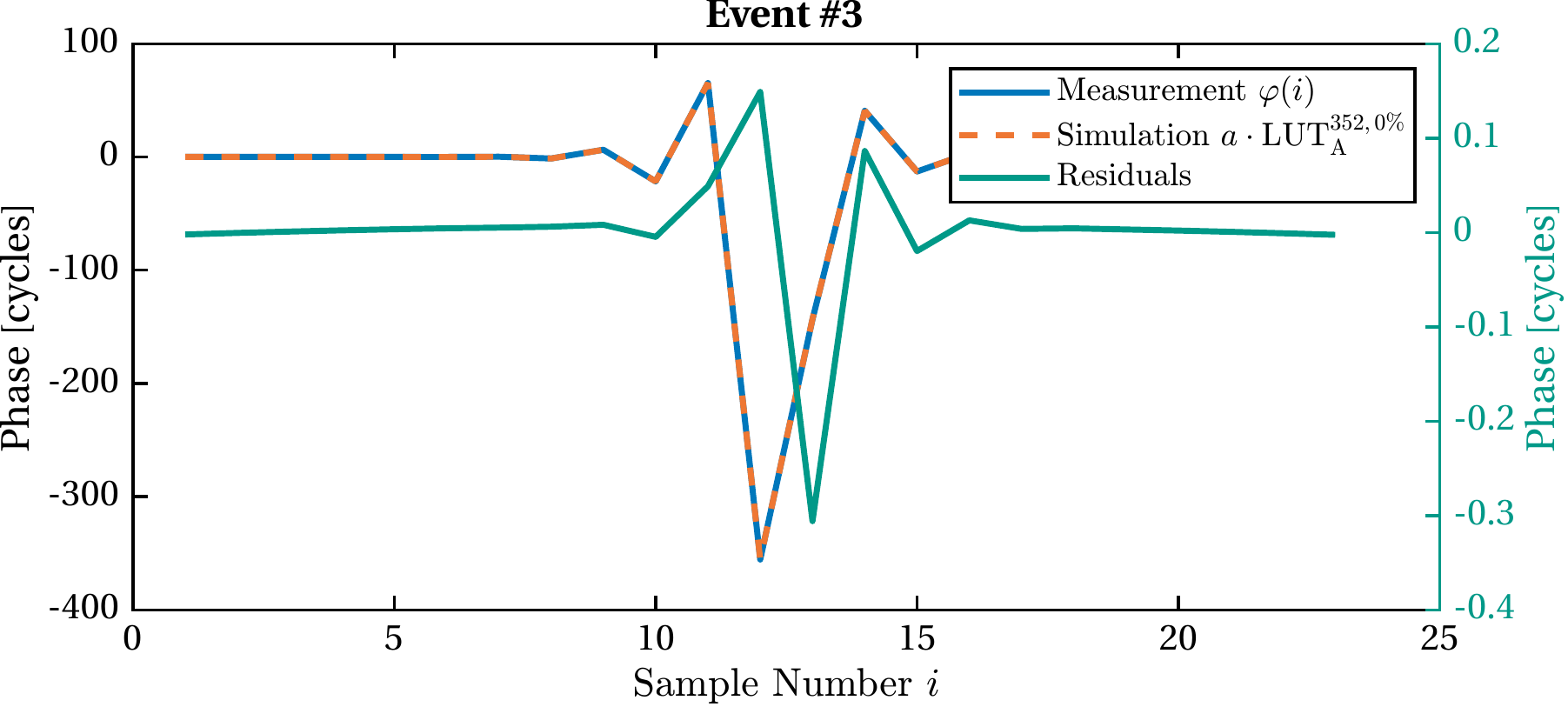}
    \caption{Event \#3: The residuals are shaped like a second \gls{SEU}.}
    \label{fig::seu_2}
\end{figure}

There are two events where the residuals still show a comparatively large rms value (\#3 and \#9; marked gray). Their residuals look like another \gls{SEU} event, separated from the initial one by a few milliseconds. Event \#3 is exemplarily shown in \cref{fig::seu_2}.
A short experiment of feeding these residuals again into the fitting algorithm did not succeed, likely because these two events lived simultaneously within the filter and can not be treated as a simple superposition of two independent events. Extending the \glspl{LUT} also to incorporate such events was beyond the scope of this study and would exponentially increase the size of the \glspl{LUT} and the computation time.

\section{Conclusion}
\label{sec::conclusion}
In this paper, we presented an approach to identify, extract and model \gls{SEU}-induced disturbances in the measured phase data of the \gls{LRI} in the \gls{GFO} mission. We explain the filtering within the \gls{LRP}, where we expect \glspl{SEU} to show an effect in the measured phase through flipped bits in the registers of the lowpass \gls{FIR} filters. Further, we showed simulated data and discussed the parameters needed to describe the \gls{SEU}. Ultimately, we found 29 events in more than three years of \gls{LRI} ranging data. The events clustered at the \glsentrylong{SAA}. Some of the events seem to originate from multiple bits flipping simultaneously or possibly even with a slight time delay.

Radiation-induced \gls{SEU} in the \gls{LRI} phase data are rare and short-lived events. 
Thus, we expect that their non-removal has only little to none impact on the retrieved gravity fields. Nevertheless, \gls{LRI} data products with removed \glspl{SEU} can be found at \url{https://www.aei.mpg.de/grace-fo-ranging-datasets}. 

This study shows that it is possible to identify and remove this particular noise source in post-processing. Future instruments might overcome this source of short measurement disturbances by implementing radiation-hardened memory or incorporating error correction algorithms.

\section*{Funding}
This work has been supported by: The Deutsche Forschungsgemeinschaft (DFG, German Research Foundation, Project-ID 434617780, SFB 1464); Clusters of Excellence ``QuantumFrontiers: Light and Matter at the Quantum Frontier: Foundations and Applications in Metrology'' (EXC-2123, project number: 390837967); the European Space Agency in the framework of Next Generation Geodesy Mission development and ESA's third-party mission support for GRACE-FO (RFP/3-17121/21/I-DT-lr); the Max Planck Society (MPG) for future mission support (M.IF.A.QOP18108) and in the framework of the LEGACY cooperation on low-frequency gravitational-wave astronomy (M.IF.A.QOP18098).

\section*{Acknowledgments}
The authors would like to thank the LRI team at JPL for helpful regular discussions and insights.

\bibliographystyle{jasr-model5-names}
\biboptions{authoryear}
\bibliography{refs}

\end{document}